\documentclass[12pt]{article}

\usepackage{amsmath}
\usepackage{graphicx}
\usepackage{eufrak}
\usepackage{cite}

\renewcommand{\theequation}{\arabic{section}.\arabic{equation}}
\newcommand{\mys}[1]{\section{#1} \setcounter{equation}{0}}

\newcommand{\myappendix}{\appendix
   \renewcommand{\theequation}{\Alph{section}.\arabic{equation}}
   \vspace{30pt} \noindent {\Large \bf Appendices}}

\newlength{\dummysp}
\newcommand{\spc}{\hbox{\hspace{\dummysp}}}
\newcommand{\diag}{\mathop{{\hbox{diag} \, }}\nolimits}
\newcommand{\tr}{\mathop{{\hbox{Tr} \, }}\nolimits}

\newcommand{\adj}{\mathop{{\hbox{{\bf adj}} \, }}\nolimits}

\newcommand{\bbar}[1]{{\overline{#1}}}
\newcommand{\half}{\frac{1}{2}}

\newcommand{\beq}{\begin{eqnarray}}
\newcommand{\eeq}{\end{eqnarray}}
\newcommand{\nnn}{ \nonumber \\ }

\newcommand{\chib}{{\bar \chi}}
\newcommand{\e}{{\epsilon}}
\newcommand{\s}{{\sigma}}
\newcommand{\vev}[1]{{\langle #1 \rangle}}

\newcommand{\gappeq}{\mathrel{\rlap {\raise.5ex\hbox{$>$}}
{\lower.5ex\hbox{$\sim$}}}}
\newcommand{\lappeq}{\mathrel{\rlap{\raise.5ex\hbox{$<$}}
{\lower.5ex\hbox{$\sim$}}}}
\newcommand{\myref}[1]{(\ref{#1})}

\newcommand{\ben}{\begin{enumerate}}
\newcommand{\een}{\end{enumerate}}

\newcommand{\sqtw}{\sqrt{2}}

\newcommand{\sbar}{{\bar \s}}

\newcommand{\hc}{{\rm h.c.}}

\newcommand{\psib}{{\bar \psi}}

\newcommand{\bit}{\begin{itemize}}
\newcommand{\eit}{\end{itemize}}

\newcommand{\obf}{{\bf 1}}

\newcommand{\mbf}{{\bf m}}

\newcommand{\Ncal}{{\cal N}}

\newcommand{\zdag}{z^\dagger}
\newcommand{\zb}{\zdag}

\newcommand{\Qt}{{\tilde Q}}
\newcommand{\Phit}{{\tilde \Phi}}
\newcommand{\Vt}{{\tilde V}}
\newcommand{\gtil}{{\tilde g}}
\newcommand{\Scal}{{\cal S}}
\newcommand{\Ccal}{{\cal C}}
\newcommand{\Acal}{{\cal A}}

\newcommand{\Ft}{{\tilde F}}
\newcommand{\ghat}{{\hat g}}
\newcommand{\Ufd}{{U(N_f)_{\text{diag}}}}
\newcommand{\Psib}{{\bar \Psi}}
\newcommand{\lamb}{{\bar \lambda}}
\newcommand{\Lamb}{{\bar \Lambda}}
\newcommand{\chit}{{\tilde \chi}}
\newcommand{\ehat}{{\hat e}}
\newcommand{\ihat}{\ehat_1}
\newcommand{\jhat}{\ehat_2}

\def\[{\left [}
\def\]{\right ]}
\def\({\left (}
\def\){\right )}

\begin{document}

\begin{titlepage}

\renewcommand{\thefootnote}{\fnsymbol{footnote}}

FTPI-MINN-06/14
\hfill Sept.~1, 2006

UMN-TH-2501/06
\hfill hep-lat/0605004

\vspace{0.45in}

\begin{center}
{\bf \Large A deconstruction lattice description of \\ \vskip 10pt
the D1/D5 brane world-volume gauge theory }
\end{center}

\vspace{0.15in}

\begin{center}
{\bf \large Joel Giedt\footnote{{\tt giedt@physics.umn.edu}}}
\end{center}

\vspace{0.15in}

\begin{center}
{\it Fine Theoretical Physics Institute, University of Minnesota \\
116 Church St.~S.E., Minneapolis, MN 55455 USA }
\end{center}

\vspace{0.15in}

\begin{abstract}

\end{abstract}
In this article, I generalize the deconstruction lattice formulation
of Endres and Kaplan [hep-lat/0604012]
to two-dimensional super-QCD with eight supercharges, denoted (4,4), and
bifundamental matter.
I specialize to a particularly interesting (4,4)
gauge theory, with gauge group $U(N_c) \times U(N_f)$,
and $U(N_f)$ weakly gauged.
It describes the infrared limit of
the D1/D5 brane system, which has been
studied extensively as an example of the $AdS_3/CFT_2$
correspondence.  The construction here preserves
two supercharges exactly, and has a lattice
structure quite similar to that which has
previously appeared in the deconstruction
approach; i.e., site, link and diagonal fields with both
Bose and Fermi statistics.
I remark on possible applications of
the lattice theory that would test the
$AdS_3/CFT_2$ correspondence, particularly one that
would exploit the recent worldsheet instanton analysis
of Chen and Tong [hep-th/0604090].
\end{titlepage}

\renewcommand{\thefootnote}{\arabic{footnote}}
\setcounter{footnote}{0}

\mys{Introduction}
Supersymmetric large $N_c$ gauge theory seems to afford
a window on quantum gravity, through the AdS/CFT correspondence
\cite{Maldacena:1997re,Gubser:1998bc,Witten:1998qj}.
Recent formulations of lattice supersymmetry give
some hope that we may be able to study these ideas
on the lattice.  In particular, to what extent does
the correspondence hold at intermediate $N_c$,
at finite temperature, and for non-BPS quantities?

Many promising lattice formulations of supersymmetric
field theories
occur in two dimensions (2d).\footnote{
For an extensive list of references on lattice
formulations of supersymmetric field theories,
both old and new, see \cite{Giedt:2006pd}.} 
In some cases, convincing perturbative arguments can be made that
the correct continuum limit is obtained
without fine-tuning \cite{Elitzur:1983nj,Kaplan:2002wv,
Cohen:2003xe,Cohen:2003qw,Giedt:2004qs,Kaplan:2005ta,
Sugino:2003yb,Sugino:2004qd,Sugino:2004uv,Sugino:2006uf}.  
(Another interesting approach, involving noncommutativity
on the scale of the lattice, deserves further study
of its quantum continuum limit~\cite{D'Adda:2004jb,D'Adda:2005zk}.)
In other cases, super-renormalizability implies that fine-tuning
a small set of one-loop diagrams allows one to
obtain the desired continuum limit in perturbation theory,
as in \cite{Golterman:1988ta}.  (A 3d analogue is \cite{Elliott:2005bd}.)

Broadly speaking, it is the softer ultraviolet (UV) divergences in 2d 
that generically make it easier to obtain the desired
continuum limit in perturbation theory.
Whether or not this property holds nonperturbatively
is an open question, that at this point can
only be answered empirically.
In this regard, it is important to note that 2d field
theories are more practical to study numerically;
a small computer cluster can obtain reasonably
accurate results.
For some 2d examples, Monte Carlo simulation
results have provided information on nonperturbative 
renormalization.  For example, recent simulations
of 2d supersymmetric theories that preserve a
nilpotent subalgebra seem entirely consistent
with continuum expectations \cite{Beccaria:1998vi,
Catterall:2001wx,Catterall:2001fr,Giedt:2005ae,
Catterall:2006jw,Catterall:2006sj}.
In this author's opinion, the encouraging
results in 2d suggest that it is 
time to look for interesting applications
of the lattice supersymmetry ideas that have 
been developed thus far.

A well-known
example of the AdS/CFT correspondence occurs in
the Type IIB superstring, at the intersection
of D1 and D5 branes, with four of the directions
of the D5 brane wrapped on, say, a torus $T^4$.
The IR limit of the world-volume intersection theory
is a 2d (4,4) supersymmetric gauge theory.  It
can be understood as the dimensional reduction
of a 4d $\Ncal=2$ super-QCD \cite{Ferrara:1974pu,
Fayet:1975yi,Fayet:1978ig} where $N_f$ flavors
of matter are contained in hypermultiplets,
and transform in the fundamental representation of
the $U(N_c)$ gauge group.  These flavors
are minimally coupled, so that there is
a $U(N_f)$ flavor symmetry.  In actuality,
the $U(N_f)$ symmetry is weakly gauged,
and the flavors are bifundamentals of $U(N_c)
\times U(N_f)$.

In this article, the
(2,2) supersymmetric formulation of
Endres and Kaplan (EK) \cite{Endres:2006ic} will be
generalized to (4,4) theories
that describe the world-volume gauge theory of the
D1/D5 brane system.
It will be seen that, in the gauge sector, a slight modification of
the original (4,4) pure SYM construction 
of Cohen et al.~(CKKU) \cite{Cohen:2003qw}
is required.  EK have shown, in general terms, how
to construct (2,2) theories with bifundamental matter
under a quiver gauge group $U(N)^m$ (``Example 2'' at
the end of their article).  
The target theory for the D1/D5 system that
is aimed at here
is (4,4) 2d super-QCD with gauge group $U(N_c)\times U(N_f)$.
In this context, we want the flexibility to choose
$N_c$ and $N_f$ independently.  Also,
as already mentioned, the flavor group $U(N_f)$ should be
weakly gauged; that is, the corresponding gauge couplings
should satisfy $g_{N_f} \ll g_{N_c}$ for all
scales of interest.  It will be seen below that it is not difficult to modify
EK's technique to obtain a quiver of two factors, $U(N_c)$
and $U(N_f)$, with $N_c \not= N_f$.
The nontrivial task is to devise a
trick to make the $U(N_f)$ weakly gauged in the EK construction
with bifundamental matter.  The trick that is used is quite standard,
and has an interesting interpretation,
as will be described below.  With these various
generalizations of the EK formulation, the lattice
theory described here is specially tailored to describe the D1/D5
world-volume theory.  Having at hand a fully
latticized description of this theory, one
can contemplate various nonperturbative studies
that would be of interest, both numerical (Monte
Carlo simulations) and analytical (strong coupling
expansions).

I now summarize the remainder of this
article:
\bit
\item
\S\ref{summ}
{\it Summary of the target theory.}
Here I describe the continuum $U(N_c) \times U(N_f)$ (4,4)
gauge theory that is aimed at, using the language of 4d $\Ncal=1$
superfields.
\item
\S\ref{latt}
{\it Lattice construction.}
First I explain in general terms how
a (4,4) theory with gauge group $U(N_c) \times U(N_f)^n$
and bifundamental matter
is obtained.  The $U(N_f)^n$ quiver is introduced
in order to obtain a weakly coupled diagonal
subgroup, $U(N_f)_{\text{diag}}$.  It will be explained how
this can be interpreted in terms of a deconstructed third dimension.
The $U(N_c)$ gauge multiplet does not propagate
in this direction, but is stuck to the 2d subspace.
It is interesting that this mimicks what occurs in the D1/D5
system, where $U(N_c)$ gauge fields are stuck
to the $D1$ brane, and $g_{N_f} \ll g_{N_c}$
is due to ``volume suppression.''  In the 4d $\Ncal=1$ language,
the lowest components of hypermultiplets are
$SO(1,3) \times U(1)_R$ neutral and form a doublet of $SU(2)_R$.
It follows that in the conventions of CKKU,
the hypermultiplets would have fractional
$N$-ality with respect to (w.r.t.)~the $Z_N \times Z_N$ that
is used to define the lattice theory.  This
unacceptable situation calls for a minor modification
of the choice of global charges
used in the $Z_N \times Z_N$ orbifold, relative
to CKKU \cite{Cohen:2003qw}.  Finally,
I describe how to break to
$U(N_f)_{\text{diag}}$, and the conditions that must be satisfied
for the Kaluza-Klein (KK) states of the corresponding third dimension
to decouple from the $U(N_c) \times U(N_f)_{\text{diag}}$
effective theory.
\item
\S\ref{appl}
{\it Application.}
Here I describe a simple study of the
characteristics of instantons in the sector
with unit first Chern class.  The distribution
of these characterisitics---instanton size
and orientation---have been shown recently
to have $AdS_3 \times S^3$ geometry \cite{Chen:2006ps}.
This support for the $AdS_3/CFT_2$ correspondence
would be interesting to study on the lattice
where we can access intermediate $N_c$ and
finite temperature.
In this regard, the recent results
of Rey and Hikida \cite{Rey:2006bz} provide continuum
results that could be compared to.
\item
\S\ref{conc}
{\it Conclusions.}
Here I summarize this work and outline
research that is in progress.
\item
{\it Appendices.}  Some technical
details and lengthy formulae have been
separated from the main text.
\eit

The purpose of this article is to give a brief
outline of the lattice construction and its
potential applications.  A more thorough discussion
of details associated with the lattice 
system (superspace description, renormalization,
etc.), as well as intensive studies of
the possible applications mentioned in \S\ref{appl}, 
are left to future work.

\mys{Summary of the target theory}
\label{summ}
The 2d theory is most easily obtained from
a dimensional reduction of the 4d theory written
in $\Ncal=1$ superspace.\footnote{See \cite{Wess:1992cp}
for a review of this formalism.}
The $U(N_c)$ $\Ncal=2$ vector multiplet is written in terms
of an $\Ncal=1$ vector superfield $V$ and an
adjoint $\Ncal=1$ chiral superfield $\Phi$.  For $U(N_f)$
the notation will be $\Vt,\Phit$.  The
action is compactly described in terms of the
(real) K\"ahler potential $K$ and the (holomorphic)
superpotential $W$.

For the gauge multiplet we have:
\beq
K_{\text{gge}} &=& \frac{1}{g^2} \Phi^{\dagger A} (e^V)_A^{\spc B} \Phi_B
+ \frac{1}{\gtil^2} \Phit^{\dagger M} (e^\Vt)_M^{\spc N} \Phit_N, \nnn
W_{\text{gge}} &=& \frac{1}{4 g^2} W^{\alpha}_A W^A_\alpha
+ \frac{1}{4 \gtil^2} \tilde W^{\alpha}_M \tilde W^M_\alpha,
\label{kwg}
\eeq
written in terms of the usual chiral field strength spinor
superfields $W^\alpha(V)$ and $\tilde W^\alpha(\tilde V)$,
and adjoint representation matrices $t^{A \spc B}_{\spc C}$
and $t^{L \spc M}_{\spc N}$, for $U(N_c)$ and $U(N_f)$
respectively.

The hypermultiplet is written in terms of two chiral multiplets,
denoted $Q$ and $\Qt$.  The $U(N_c) \times U(N_f)$ representations
for these superfields are:
\beq
Q_a^{\spc m} = (N_c,\bar N_f), \quad
\Qt_m^{\spc a} = (\bar N_c, N_f).
\label{uyr}
\eeq
The indices range according to:  $a=1,\ldots,N_c; \; m=1,\ldots,N_f$.
It will be convenient to regard $Q$ as an $N_c \times N_f$
matrix, and $\Qt$ and an $N_f \times N_c$ matrix.  Correspondingly,
$Q^\dagger$ will be an $N_f \times N_c$ matrix and
$\Qt^\dagger$ will be an $N_c \times N_f$ matrix.
We can then write the K\"ahler potential as:
\beq
K_{\text{mat}} &=& \tr Q^\dagger e^V Q e^{-\Vt} 
+ \tr \Qt^\dagger e^\Vt \Qt e^{-V} \nnn
&=& (Q^\dagger)_m^{\spc a} (e^V)_a^{\spc b} Q_b^{\spc n} (e^{-\Vt})_n^{\spc m}
+ (\Qt^\dagger)_a^{\spc m} (e^\Vt)_m^{\spc n} \Qt_n^{\spc b} (e^{-V})_b^{\spc a}.
\label{mmkp}
\eeq
Note that $V$ is expressed in terms of $U(N_c)$
fundamental representation generators $t^{A \spc b}_{\spc a}$;
a similar statement holds for $\Vt$, except that
the group is $U(N_f)$.
The normalization convention that is assumed in
the following is defined by $\tr t^A t^B = (1/2) \delta^{AB}$
for the fundamental representation.
In the second step of \myref{mmkp}, the indices have been written explicitly
in order to make the matrix notation clear.  Below,
such details will be left implicit.  

The superpotential is the minimal one, which
preserves $U(1)_R$:
\beq
W_{\text{mat}} = \sqtw \tr \Qt \Phi Q - \sqtw \tr Q \Phit \Qt.
\eeq
Here, $\Phi$ and $\Phit$ are expressed in terms of $U(N_c$)
and $U(N_f)$ fundamental representation generators respectively.
It is easy to check gauge invariance, which
acts holomorphically on the chiral superfields:
\beq
&& Q \to e^\Lambda Q e^{-\tilde \Lambda}, \quad
\Qt \to e^{\tilde \Lambda} \Qt e^{-\Lambda}, \quad
e^V \to e^{-\Lambda^\dagger} e^V e^{-\Lambda}, \quad
e^\Vt \to e^{-\tilde \Lambda^\dagger} e^\Vt e^{-\tilde \Lambda},
\nnn && \Phi \to e^\Lambda \Phi e^{-\Lambda}, \quad
\Phit \to e^{\tilde\Lambda} \Phit e^{-\tilde\Lambda},
\eeq
where $\Lambda$ and $\tilde\Lambda$ are chiral
superfields valued in the Lie algebras of $U(N_c)$
and $U(N_f)$ respectively.

The action is given by a Grassmann integral
over superspace coordinates $\theta^\alpha,
\bar \theta_{\dot \alpha}$:
\beq
S = \int d^4x \bigg\{ \int d^4\theta \( K_{\text{gge}} +
K_{\text{mat}} \) + \[ \int d^2\theta \( W_{\text{gge}} +
W_{\text{mat}} \) + \text{h.c.} \] \bigg\}.
\eeq

\mys{Lattice construction}
\label{latt}
The EK approach \cite{Endres:2006ic} includes matter, in a generalization of 
earlier work by Kaplan et al., especially 
CKKU~\cite{Kaplan:2002wv,Cohen:2003xe,Cohen:2003qw,
Kaplan:2005ta}.  The Kaplan et al.~``deconstruction
lattice'' approach was an outgrowth of
{\it dimensional deconstruction} 
\cite{Arkani-Hamed:2001ca,Hill:2000mu}.
In ``Example 2'' given by EK, quiver gauge
theories with bifundamental matter were formulated.
In this section, I generalize EK's quiver
construction to the case of (4,4)
2d super-QCD with bifundamental matter
that is charged under a gauge group
$U(N_c) \times U(N_f)$.  A minor modification of the
(4,4) setup of CKKU \cite{Cohen:2003qw} will prove necessary,
due to the R-charges of the hypermultiplets that are
being added to the theory.  The other difficulty will
be that we need to have $U(N_f)$ weakly gauged relative to $U(N_c)$.
This will be addressed through extending to a quiver gauge
theory $U(N_c) \times U(N_f)^n$, and then breaking
$U(N_f)^n$ to its diagonal subgroup.

\subsection{Outline}
In the present theory, we begin with a matrix model
that is the zero-dimensional (0d) reduction\footnote{
The 0d reduction is obtained by treating all
fields as independent of space-time coordinates.}
of 4d $\Ncal=2$ super-QCD with gauge group $U((N_c + n N_f)N^2)$
and fundamental matter.
This theory is described by:
\beq
&& K_{\text{gge}} = \frac{1}{g^2} \Phi^{\dagger A} 
(e^V)_A^{\spc B} \Phi_B, \quad
W_{\text{gge}} = \frac{1}{4 g^2} W^{\alpha}_A W^A_\alpha,
\nnn
&&
K_{\text{mat}} = Q^{\dagger a} (e^V)_a^{\spc b} Q_b 
+ \Qt^a (e^{-V})_a^{\spc b} \Qt^\dagger_b, \quad
W_{\text{mat}} = \sqtw \Qt^a \Phi_a^{\spc b} Q_b.
\label{mta}
\eeq
Here, indices $A,B$ correspond to the adjoint representation, whereas
the indices $a,b$ correspond to the fundamental representation.
We will absorb the overall space-time volume $V_4 = \int d^4x$
associated with the 4d $\to$ 0d reduction into a
redefinition of the coupling constant $g^2$ and the
matter fields $Q,\Qt$.  The resulting 0d theory (fixed to
Wess-Zumino gauge) will be referred to as the {\it mother theory,}
following Kaplan et al.

The next step is to perform an orbifold projection
on the mother theory, in order to reduce it to
the {\it daughter theory.}  This ``orbifolding'' proceeds in two steps.
First we orbifold by a $Z_{n+1}$ symmetry of the mother
theory, to break the gauge group according to:
\beq
U((N_c + n N_f)N^2) \to U(N_c N^2) \times U(N_f N^2)^n.
\label{nsc}
\eeq
Then we orbifold by a $Z_N \times Z_N$ symmetry of the
mother theory to break the gauge group further,
according to:
\beq
U(N_c N^2) \times U(N_f N^2)^n \to
U(N_c)^{N^2} \times U(N_f)^{nN^2}.
\label{osoc}
\eeq

It is at this point that the trick to get a weakly gauged
$U(N_f)$ comes in.  At the final stage of the
orbifolding---the r.h.s.~of~\myref{osoc}---the gauge coupling is universal, with its
strength determined by the single coupling $g^2$
that appears in the original 0d matrix model \myref{mta}
and the overall lattice spacing that
is determined by the choice of vacuum (dynamical
lattice spacing) for the
deconstruction---what was called the {\it $a$-configuration} 
in \cite{Giedt:2006pd}.
However, we now ``Higgs'' the subgroup $U(N_f)^{nN^2}$
on the r.h.s.~of \myref{osoc}
with universal vacuum expectation values in bifundamental
matter of this group, to break to the diagonal subgroup:
\beq
U(N_f)^{nN^2} \to U(N_f)^{N^2}_{\text{diag}}.
\eeq
Then the coupling for the diagonal group is:
\beq
\gtil^2_{\text{diag}} = g^2 /n.
\label{geff}
\eeq
For large $n$ we obtain the desired result---a weakly
gauged flavor group.

An alternative picture of this trick is the
following.  We may regard the factor $n$ as
counting sites in a third dimension that
has been deconstructed.  Only the fields
with $U(N_f)$ charge propagate in this third
dimension.  The $U(N_c)$ vector multiplet
is stuck to the 2d subspace.  It is very
interesting that this mimicks what happens
in the D1/D5 brane system.  There, the flavored
fields propagate throughout the torus $T^4$, since
they correspond to strings that have
one end on the D5 brane that wraps $T^4$.
The D1 branes are stuck at a point in
$T^4$, and so the purely colored fields
do not propagate in the $T^4$ direction.
The difference here is that, to simplify
the lattice construction, we have only a
line interval in the extra dimension.  It would
be interesting to generalize the present
construction to a $U((N_c+n^4N_f)N^2)$ mother
theory and to make a deconstructed $T^4$
appear in the theory.\footnote{The more
exotic case of a K3 manifold in the
extra four dimensions could also be attempted.}

From this perspective
we see that it is necessary to keep the
third dimension small so that we never see the
effects of the KK states.  That is, we 
want only the $U(N_f)^{N^2}_{\text{diag}}$ states
to be light enough to play a role at the scales
that we study.  In fact, this is exactly what
happens in the D1/D5 system.  Dimensional
reduction of the D5 theory to the 2d intersection
gives a volume suppression:
\beq
g^2_{\text{D5 reduc.}}/g^2_{D1} \approx \ell_s^4/V_4
\eeq
where $V_4$ is the volume of the torus $T^4$ and
$\ell_s$ is the string length.  For $V_4 \gg \ell_s^4$,
the 2d $U(N_f)$ is weakly gauged, and the KK
states are super-massive on the scale\footnote{Recall
that in 2d, $[g_{\text{D1}}]=1$, and that this is
the scale of non-KK modes.} $g_{\text{D1}}$.

In the discussion of \S\ref{yyu} below, details associated
with decoupling the KK states along the third
dimension will be addressed.

\subsection{Mother theory}
In $\Ncal=1$ superfield notation, the mother theory
is the 0d reduction of \myref{mta}.
It is straightforward to work out the 0d reduction
of the component field action in the mother theory.
I denote component fields (in Wess-Zumino gauge):
\beq
V=(v_\mu,\lambda,\lamb,D), \quad
\Phi=(\phi,\psi,G), \quad Q=(Q,\chi,F),
\quad \Qt=(\Qt,\chit,\Ft).
\eeq
The result, after Euclideanization, is:
\beq
S_{gge} &=& \frac{1}{2g^2} \tr [v_\mu,v_\nu] [v_\mu,v_\nu]
+ \frac{2}{g^2} \tr [v_\mu,\phi][v_\mu,\phi^\dagger] \nnn
&& + \frac{1}{g^2} \tr \( D^2 + 2 D [\phi,\phi^\dagger] \)
+ \frac{2}{g^2} \tr G^\dagger G \nnn
&& + \frac{2}{g^2} \tr \Psib [v_\mu \gamma_\mu, \Psi]
+ \frac{2\sqtw i}{g^2} \tr \( \lambda [\psi,\phi^\dagger]
- [\phi,\psib] \lamb \), 
\label{gmom} \\
S_{mat} &=& -Q^\dagger v_\mu v_\mu Q + F^\dagger F + Q^\dagger D Q
- \Qt v_\mu v_\mu \Qt^\dagger + \Ft \Ft^\dagger \nnn
&& - \Qt D \Qt^\dagger 
+ \sqtw \( \Ft \phi Q + \Qt G Q + \Qt \phi F + \hc \) \nnn
&& + \Lamb v_\mu \gamma_\mu \Lambda 
- \sqtw \( \chit \phi \chi + \chib \phi^\dagger \bar \chit
+ \chit \psi Q + \Qt \psi \chi + Q^\dagger \psib \bar \chit
+ \chib \psib \Qt^\dagger \) \nnn
&& + i \sqtw \( Q^\dagger \lambda \chi - \chib \lamb Q
- \chit \lambda \Qt^\dagger + \Qt \lamb \bar \chit \).
\label{mmom}
\eeq
Here, the following notations are used ($\alpha=1,2$):
\beq
&& \Psi = \binom{\lambda_\alpha}{\psib^{\dot \alpha}}, 
\quad \Psib = (\psi^\alpha,\lamb_{\dot \alpha}),
\quad \Lambda = \binom{\chi_\alpha}{\bar \chit^{\dot \alpha}}, \quad
\Lamb = (\chit^\alpha,\chib_{\dot \alpha}), \nnn
&& \gamma_\mu = \begin{pmatrix} 0 & \s_\mu \cr \sbar_\mu & 0
\end{pmatrix}, \quad \s_\mu = (\vec \s,i), \quad 
\sbar_\mu = (-\vec \s,i), \quad \tr T^A T^B = \half \delta^{AB}.
\label{lodf}
\eeq
It is not difficult to relate \myref{gmom} to the
mother theory action of CKKU.  The translation is:
\beq
&& z_1 = \frac{1}{\sqtw}(v_1+iv_2), \quad
z_2 = \frac{1}{\sqtw}(v_3+iv_4), \quad z_3 = i \phi^\dagger, \nnn
&& \Psi = (\xi_2,\xi_1,\xi_3,\lambda), \quad
\Psib = (\psi_1, -\psi_2, \chi, - \psi_3).
\eeq
Note that $\lambda,\chi$ here are not the two-component
fermions $\lambda_\alpha,\chi_\alpha$ of the 4d notation \myref{lodf}.
The $U(1)^4$ subgroup of $SO(6) \times SU(2)_R$ that
CKKU choose for their orbifold procedure is then
\beq
&& q_1 = \Sigma_{1,2} = \frac{1}{4i} [\gamma_1,\gamma_2], \quad
q_2 = \Sigma_{3,4} = \frac{1}{4i} [\gamma_3,\gamma_4], \nnn
&& q_3 = -\half Q_R, \quad q_4 = -T^3_R.
\eeq
I have expressed the last two charges in terms of the conventional
$SU(2)_R \times U(1)_R$ R-symmetry of the 4d $\Ncal=2$ theory.
The $U(1)_R$ generator $Q_R$ is normalized such that gluinos (denoted $\lambda_\alpha,
\psib^{\dot \alpha}$ in \myref{gmom}-\myref{lodf}) have
$Q_R=1$.  $T^3_R = (1/2)\s^3$ is the diagonal generator
of $SU(2)_R$.  The charges of all gauge multiplet fields are
summarized in Table \ref{vmr}.  The charges of all hypermultiplet
fields are summarized in Table \ref{mtab}.
For the fermions, the notation is related to Eqs.~\myref{gmom}-\myref{lodf}
by
\beq
(\Psi_1, \Psi_2, \Psi_3, \Psi_4) = (\lambda_1, \lambda_2,
\psib^{\dot 1} \psib^{\dot 2}), \quad
(\Psib_1,\Psib_2,\Psib_3,\Psib_4) = (\psi^1, \psi^2, \lamb_{\dot 1}
\lamb_{\dot 2}),
\eeq
and a similar translation for $\Lambda, \Lamb$.
The upper and lower placement of indices is significant
because of the implicit spinor sums that are
in \myref{gmom}-\myref{mmom}.  For example, in
the last line of \myref{mmom}, one has the term
\beq
Q^\dagger \lambda \chi = Q^\dagger \lambda^\alpha \chi_\alpha
= Q^\dagger \( \e^{12} \lambda_2 \chi_1 + \e^{21} \lambda_1 \chi_2 \)
= Q^\dagger \( \Psi_2 \Lambda_1 - \Psi_1 \Lambda_2 \).
\eeq
Here, the conventions of \cite{Wess:1992cp} have been used:
$\e_{21}=\e^{12}=1, \; \e_{12}=\e^{21} = -1$.
These details were important in writing down
the explicit daughter theory action that is given in
Appendix \ref{dmac}.

\begin{table}
$$
\begin{array}{cccccccccccccc}
&z_1& z_2& i\phi^\dagger & \Psi_1 & \Psi_2 & \Psi_3 & \Psi_4 & \Psib_1 & \Psib_2 & \Psib_3 
& \Psib_4 & G & D \\
& & & z_3 & \xi_2 & \xi_1 & \xi_3 & \lambda & \psi_1 & -\psi_2 & \chi & -\psi_3 
& & \\ \hline
q_1 & 1 & 0 & 0 & -\half & +\half & -\half & +\half & +\half & -\half 
& +\half & -\half & 0 &0 \\
q_2&0&1&0 & +\half & -\half & -\half & +\half& -\half& +\half& +\half& -\half&0&0 \\
q_3&0&0&1 & -\half & -\half & +\half & +\half& -\half& -\half& +\half& +\half&0&0 \\
q_4&0&0&0 & -\half & -\half & -\half & -\half& +\half& +\half& +\half& +\half& -1 &0 \\ \hline
r_1 & 1 & 0 & -1 & 0 & 1 & -1 & 0 & 1 & 0 & 0 & -1 & 0 & 0 \\ 
r_2 & 0 & 1 & -1 & 1 & 0 & -1 & 0 & 0 & 1 & 0 & -1 & 0 & 0 \\ \hline
\end{array}
$$
\caption{$U(1)^4$ charges and $Z_N \times Z_N$
orbifold action $N$-alities for the gauge multiplet, after modification \myref{mmod}
to accomodate hypermulitiplets.  The second
line connects to the CKKU notation for the fields
$z_3, \xi_2, \xi_1$, etc.
\label{vmr}}
\end{table}

\begin{table}
$$
\begin{array}{ccccccccccccc}
&Q & \Qt & \Lambda_1 & \Lambda_2 & \Lambda_3 & \Lambda_4 & \Lamb_1 
& \Lamb_2 & \Lamb_3 & \Lamb_4 & F & \Ft \\ \hline
q_1 & 0 & 0 & -\half & +\half & -\half & +\half & +\half & -\half 
& +\half & -\half & 0 & 0 \\
q_2 & 0 & 0 & +\half & -\half & -\half & +\half& -\half& +\half& +\half& -\half
& 0 &0 \\
q_3 & 0 & 0&+\half&+\half& -\half& -\half &+\half&+\half& -\half& -\half&1&1\\
q_4 & +\half & +\half&0&0&0&0&0&0&0&0&-\half&-\half \\ \hline
r_1 & 0 & 0 & -1 &  0 & 0 & 1 & 0  & -1 & 1 & 0 & -1 & -1 \\
r_2 & 0 & 0 &  0 & -1 & 0 & 1 & -1 &  0 & 1 & 0 & -1 & -1 \\ \hline
\end{array}
$$
\caption{$U(1)^4$ charges and $Z_N \times Z_N$
orbifold action $N$-alities for the matter hypermultiplets. 
\label{mtab}}
\end{table}

\subsection{Orbifolding details}

\subsubsection{Projections, generally}
Denote $U((N_c + n N_f)N^2)$ indices collectively
\beq
\Scal \equiv I m_1 m_2 , \quad
I \in \{1,\ldots,N_c + n N_f \}, \quad
m_1,m_2 \in \{ 0,\ldots,N \} .
\label{scmm}
\eeq
The domain of the index $I$ should be thought of
as follows:
\beq
&& I = 1,\ldots,N_c;N_c+1,\ldots,N_c+N_f;(N_c+N_f)+1,\ldots,(N_c+N_f)+N_f;\ldots 
\nnn && \ldots;(N_c+(n-1)N_f)+1,\ldots,(N_c+(n-1)N_f)+N_f.
\eeq
The interpretation is in terms of a block diagonal
matrix, with an $N_c \times N_c$ block, followed
by $n$ blocks of size $N_f \times N_f$.  The index $\Scal$
then indicates, say, that the entries of the $N_c \times N_c$
matrix are themselves $N^2 \times N^2$ matrices, and so on.
In what follows, ``diag'' will indicate a block diagonal
matrix, with only block entries given explicitly.  For
example the unit matrix in the mother theory is given by
\beq
\obf_{(N_c + n N_f)N^2} = \diag(\obf_{N_cN^2}, 
\obf_{N_fN^2}, \ldots, \obf_{N_fN^2}),
\eeq
with $n$ entries of $\obf_{N_fN^2}$.  Other matrices
of this form follow.

Introduce ``clock operators'' that involve roots of unity
$\omega_k \equiv \exp(2\pi i/k)$:
\beq
&& P = \diag( \obf_{N_cN^2}, 
\omega_{n+1} \obf_{N_fN^2}, \ldots, \omega_{n+1}^n \obf_{N_fN^2} ), 
\nnn && \Omega_N = \diag(1,\omega_N,\ldots,\omega_N^{N-1}), \nnn
&& C^{1,k}_N = \obf_{k} \otimes \Omega_N \otimes \obf_N, \quad
C^{2,k}_N = \obf_{k} \otimes \obf_N \otimes \Omega_N, \nnn
&& \Ccal^1_N = \diag(C_N^{1,N_c}, C_N^{1,N_f},\ldots, C_N^{1,N_f}), \nnn
&& \Ccal^2_N = \diag(C_N^{2,N_c}, C_N^{2,N_f},\ldots, C_N^{2,N_f}).
\eeq
Orbifold projections for any field $\Acal$ are defined by:
\beq
\Acal \equiv \omega_{n+1}^s P \Acal P^\dagger, \quad
\Acal \equiv \omega_{N}^{r_1} \Ccal^1_N \Acal \Ccal^{1 \dagger}_N, \quad
\Acal \equiv \omega_{N}^{r_2} \Ccal^2_N \Acal \Ccal^{2 \dagger}_N.
\label{prsu}
\eeq
The charges $s,r_1,r_2$ will correspond to, respectively, $(n+1)$-ality,
$N$-ality, $N$-ality.  The origin of the $Z_{n+1}$ symmetry
in the mother theory---corresponding
to the $(n+1)$-ality---will be discussed shortly.
The two $N$-alities correspond to
a $Z_N \times Z_N$ subgroup of the $SO(6) \times SU(2)_R$
symmetry of the mother theory.

To understand the effect of \myref{prsu}, it is best to look 
at it in stages.
For the $Z_{n+1}$ projection,
\beq
\Acal_{s=0} &\to& (\adj U(N_cN^2),1,\ldots,1) \oplus (1,\adj U(N_fN^2),1,
\ldots,1) \oplus \cdots \nnn
&& \oplus (1,1,\ldots,\adj U(N_fN^2)) \nnn
\Acal_{s=1} &\to& (N_cN^2,\bbar{N_fN^2},1,\ldots,1) \oplus
(1,N_fN^2,\bbar{N_fN^2},1,\ldots,1) \oplus \cdots \nnn
&& \oplus (1,1,\ldots,1,N_fN^2,\bbar{N_fN^2}) \oplus
(\bbar{N_cN^2},1,\ldots,1,N_fN^2)
\label{hury}
\eeq
and $s= -1$ is conjugate to the latter.
This yields ``sites'' and ``links'' of
the quiver gauge theory \myref{nsc}.
This structure will persist in the lattice
theory and its continuum limit.

The minimal coupling superpotential of the
mother theory has the $U(1)$ global symmetry
\beq
Q \to e^{i\alpha} Q, \quad \Qt \to e^{-i\alpha} \Qt,
\label{flas}
\eeq
with all other fields neutral and all components
in $Q,\Qt$ transforming identically (it is not
an R-symmetry).  This is the symmetry that we
use for the $Z_{n+1}$ orbifold.  That is,
we assign $s=1$ to all components of $Q$,
$s= -1$ to all components of $\Qt$, and
$s=0$ to all components of $V,\Phi$.
In this way, $Q,\Qt$ will be bifundamentals (``links'')
of $U(N_c) \times U(N_f)^n$ quiver gauge theory,
whereas $V,\Phi$ will be adjoints (``sites'').
The notion of ``links'' and ``sites'' used
here is distinct from that associated with the
2d lattice that is described next.
Note that because \myref{flas} is not an
R-symmetry (i.e., it commutes with the
supercharges of the mother theory), the
$Z_{n+1}$ orbifold projection leaves all
eight supercharges intact.

The $Z_N \times Z_N$ projections involve clock
matrices $\Ccal^{1,2}_N$, which only act
on the ``site'' indices $m_1,m_2$ of \myref{scmm}.
They have the usual effect of the deconstruction
lattice formulation.
Label any of the fields in the decomposition \myref{hury} by
$\Acal_{m_1,m_2;n_1,n_2}$, ignoring indices of the
surviving gauge group.
Then the surviving components of $\Acal_{m_1,m_2;n_1,n_2}$
after the $Z_N \times Z_N$
orbifold are those that satisfy:
\beq
m_1 - n_1 + r_1 = 0 \mod N; \quad m_2 - n_2 + r_2 = 0 \mod N.
\eeq
This yields site, horizontal link, vertical link,
and diagonal link interpretations,
depending on $r_1,r_2$.
The fields are then labeled by site indices $m \equiv (m_1,m_2)$.
Next I discuss particulars with respect to the various fields of the
mother theory.

\subsubsection{Daughter theory gauge action}
A problem arises for the construction of CKKU \cite{Cohen:2003qw} when we
include hypermultiplets.  The scalar components are neutral w.r.t.~the
$SO(6)$ global symmetry of the mother theory, which decomposes
to $SO(4) \times U(1)_R$ in the 4d theory.  In the
notation of CKKU, $q_1=q_2=q_3=0$.  On the other hand
these scalars transform as doublets $(\Qt^\dagger,Q)$ under $SU(2)_R$,
and as a consequence $q_4 = 1/2$ for $Q,\Qt$.
The $N$-alities defined by CKKU are
\beq
r_1 = q_1 + q_4, \quad r_2 = q_2 + q_4,
\label{hier}
\eeq
which would lead to half-integral $r_1,r_2$ for
the scalars $Q,\Qt$.
Nonintegral $N$-alities do not make sense
in the lattice interpretation of the orbifolded theory.
Therefore we must modify the $N$-ality assignments
of CKKU in order to include hypermultiplets.  It
will be seen that this is easily
accomplished.  The lattice that is obtained
is quite similar to the one of CKKU.  It is of particular
importance that two supercharges are preserved exactly.

In the modification, we want to leave the
$N$-alities of link bosons $z_1,z_2$ unchanged,
since these must ultimately get a vacuum expectation
value that links neighboring sites.\footnote{Actually,
it is an interesting question whether or
not a dynamical lattice spacing can be
associated with, say, diagonal link bosons.
I will not pursue this here.}  
We must choose $r_1,r_2$ such that all fields have
integer $N$-ality.  Also we would like to
preserve two supercharges, as in the pure gauge construction
of CKKU.  According to the CKKU rubric, we therefore
must choose $r_1,r_2$ such that two components of
the fermions are neutral.  Here we choose to keep $\Psi_4=\lambda$
neutral, as in the CKKU construction.  (Other choices
are of course possible, but lead to similar
results, due to the symmetries of the mother 
theory.) 
Then it is easy to show (cf.~Appendix \ref{prfa}) that the unique choice 
that satisfies all of our requirements is:
\beq
r_1 = q_1 - q_3, \quad r_2 = q_2 - q_3.
\label{mmod}
\eeq
In addition to $\Psi_4$, the fermion
component $\Psib_3=\chi$ is $r_1,r_2$ neutral.
The charges for the
vector multiplet are summarized in Table \ref{vmr}.

Relative to the formulation of CKKU, only the following
five fields of the gauge multiplet change their nature:
\beq
&& z_3 : \text{site} \to  (-)~ \text{diagonal link}, \nnn
&& \xi_2: (-\ihat)~ \text{link} \to \jhat~ \text{link}, \nnn
&& \xi_1: (-\jhat)~ \text{link} \to \ihat~ \text{link}, \nnn
&& \chi: \text{diagonal link} \to \text{site}, \nnn
&& \psi_3: \text{site} \to (-)~ \text{diagonal link}.
\eeq
This merely leads to modest changes in the site labels
for the daughter theory action of CKKU, their eqs.~(1.2) and (1.4).
These changes are all obvious from the $r_1,r_2$ assignments
of Table \ref{vmr}.  For instance, in their bosonic action
one replaces:\footnote{CKKU use the notation $\bar z$ where
$\zdag$ is used here.}
\beq
[\zb_{3,m}, z_{3,m}] \to
\zdag_{3,m} z_{3,m+\ihat+\jhat} - z_{3,m} \zdag_{3,m-\ihat-\jhat},
\eeq
to take into account that $z_3,\zb_3$ are now
$-/+$ diagonal link fields.
The fermion action is still of the form
\beq
S_{F,g} = \frac{2\sqtw}{g^2} \sum_m \tr \bigg\{
\Delta_m(\lambda,\zb_a,\psi_a) - \Delta_m(\chi,\zb_a,\xi_a)
+ \e_{abc} \Delta_m(\psi_a,z_b,\xi_c) \bigg\},
\eeq
with $\Delta(A,B,C)=ABC-ACB$ and site labels
assigned according to the nature of the fields
that appear.

Refering to Table \ref{vmr}, we note that
there is a two-fold degeneracy for the $r_1,r_2$
charges among the fermions.  The reason for this
is that the orbifold charges \myref{mmod}
do not involve the $SU(2)_R$ diagonal generator
$q_4$.  Thus the $SU(2)_R$ symmetry of the
mother theory is preserved, unlike that which occurs
in the CKKU construction.  Since the (4,4) gluinos
fall into doublets of $SU(2)_R$, we are
guaranteed to have the two-fold degeneracy
w.r.t.~$r_1,r_2$.

Note also that the $r_1,r_2$ neutral fermions
are those that have $q_1=q_2=q_3$.  It follows
that the two supercharges that are preserved
in the daughter theory are those that have
$q_1=q_2=q_3$.

\subsubsection{Daughter theory matter action}
Having explained how the gauge action is modified,
we next turn to the matter action.  The daughter theory
is obtained in a simple application of the orbifold
procedure to the mother theory \myref{mmom},
as determined by the
$r_1,r_2$ assignments that appear in Table~\ref{mtab}.
Due to the CKKU calculus, we are assured to obtain the
correct classical continuum limit, just
as in the EK examples.  

We have already seen from the discussion of the daughter theory gauge action that
there are two supercharges that are neutral w.r.t.~the
$Z_N \times Z_N$ charges $r_1,r_2$.  This symmetry of the 
matter mother theory action will be an exact supersymmetry of
the matter daughter theory as well.  Upon
inspection of Table \ref{mtab}, one sees
that the $r_1,r_2$ neutral fermions
are once again those that have $q_1=q_2=q_3$.  It follows
that the two supercharges that are preserved
in the daughter theory matter action are those that have
$q_1=q_2=q_3$.  It is no accident that this is
identical to what occurs in the daughter theory gauge action:
the supercharges are inherited from the mother theory.
This illustrates the usefulness of the orbifold
technique of CKKU.

Straightforward manipulations yield the
daughter theory matter action.  One merely
writes out the fermion components in \myref{mmom}
explicitly, re-expresses $v_\mu$ in terms
of $z_i,\zdag_i$, and adds site labels
as determined by the $r_1,r_2$ charges given in
Table \ref{mtab}.  Because the result
is somewhat lengthy, it has been relegated
to Appendix \ref{dmac}.

\subsection{Higgsing details}
To ``Higgs'' the theory, such that only the $U(N_f)_{\text{diag}}$
subgroup of $U(N_f)^n$
survives at the scale $g_c=ga$ of the $U(N_c)$ gauge theory,
we only require the application of the deconstruction
idea to the $U(N_f)^n$ quiver.  This 1d quiver is
similar to that considered in \cite{Hill:2000mu},
in that it is an extra dimensional interval (in this
case a third dimension),
$U(N_f)_1 \times \cdots \times U(N_f)_n$,
and it is not necessary to rework all the details.

In terms of $\Ncal=1$ superfields, the quiver
is described by the 0d reduction of the theory with
\beq
K_{mat} &=& \sum_{i=1}^{n-1} \tr \bigg\{ Q_i^\dagger e^{V_i} Q_i e^{-V_{i+1}}
+ \Qt_i^\dagger e^{V_{i+1}} \Qt_i e^{-V_i} \bigg\}, \nnn
W_{mat} &=& \sqtw \sum_{i=1}^{n-1} \tr \bigg\{ \Qt_i \Phi_i Q_i
- Q_i \Phi_{i+1} \Qt_i \bigg\}.
\eeq
Formally, this is quite similar to the quiver
theory studied in \cite{Giedt:2003xr}.
I do not write $K_{gge},W_{gge}$ since it is just an
$n$-fold replication of terms of the form \myref{mta}.
Holomorphic gauge invariance is given by
\beq
&& Q_i \to e^{\Lambda_i} Q_i e^{-\Lambda_{i+1}}, \quad
\Qt_i \to e^{\Lambda_{i+1}} \Qt_i e^{-\Lambda_{i}}, \nnn
&& \Phi_i \to e^{\Lambda_i} \Phi_i e^{-\Lambda_i}, \quad
e^{V_i} \to e^{-\Lambda_i^\dagger} e^{V_i} e^{-\Lambda_i}.
\eeq

One then gives an expectation value to the $(N_f^{(i)},\bar N_f^{(i+1)}),
\; i=1,\ldots,n-1$ bifundamentals and their conjugates:
\beq
\vev{Q_i} = \vev{\Qt_i} = \frac{1}{\sqtw a_3}.
\eeq
Then, for instance, the quadratic terms in the
2d lagrangian for the gauge bosons are\footnote{Here
I am hiding all the details of the 2d lattice theory,
and just emphasizing the quiver in the third dimension.
The modes of the lattice theory
that are getting mass here are just the
$z_{1,\mbf},z_{2,\mbf}$ that
transform as adjoints of the $U(N_f)^n$ group,
excepting the combination corresponding to
$U(N_f)_{\text{diag}}$.}
\beq
\sum_{i=1}^{n-1} \frac{g^2}{4a_3^2} (A_{i+1}^{\mu,m}-A_{i}^{\mu,m})^2,
\eeq
where a contraction over the 4d index $\mu$ and
$U(N_f)$ index $m=1,\ldots,N_f^2$ is implied.  The
scaling $A \to \ghat A$ has been performed
to make the kinetic terms for gauge bosons
canonical.  Here, $\ghat = g a^2$ is the dimensionless
coupling; i.e., the coupling of the matrix model
expressed in units of the 2d lattice.
It follows immediately from the considerations of \cite{Hill:2000mu}
that only $U(N_f)_{\text{diag}}$ has a massless
gauge boson.  All other modes are quanta with configuration
energies of order $1/(na_3)$, corresponding to discrete momenta
in the third dimension.  To be precise, the spectrum is
\beq
M_n^2 = \frac{\ghat^2}{a_3^2} \sin^2 \frac{j\pi}{n}, \quad
j=0,\ldots,n-1.
\eeq
The radius $R$ of this third deconstructed dimension
and the KK mass scale $M$ are therefore
\beq
R \approx na_3/\ghat, \quad M = \pi/R.
\eeq
The effective gauge coupling of the $U(N_f)_{\text{diag}}$
theory is given by \myref{geff}.

\subsection{Decoupling KK states}
\label{yyu}
The condition that the KK states decouple from the $U(N_c) \times \Ufd$
gauge theory is just $M \gg g_c=ga$.  Various realizations
of this could be imagined.  A strong one is that we set the KK scale
at the UV cutoff of the $U(N_c) \times \Ufd$ gauge theory: $R \equiv a$.
This translates into
\beq
n= a\ghat/a_3 = g_c \frac{a^2}{a_3}.
\eeq
Thus as we take the continuum limit $a \to 0$ in the 2d $U(N_c) \times
\Ufd$ gauge theory, with $n,g_c$ held fixed,
we have the scaling $a_3 \sim a^2$.  This
would decouple the effects of the $U(N_f)^n$ quiver at the
UV scale of the $U(N_c) \times \Ufd$ gauge theory, and just represents
a slightly different UV completion that should not have
physical consequences---based on universality arguments.

A less aggressive prescription is to take $g_cR$ fixed
and small.  This should also decouple the KK states before
important $U(N_c) \times \Ufd$ physics sets in.  This translates into
\beq
a_3/a = f/n, \quad f \ll 1.
\eeq
Holding the factors $f,n$ fixed, we see that
a scaling $a_3 \sim a$ is prescribed 
as the continuum limit is taken.

\mys{Application}
\label{appl}
Here I mention one possible application of the
lattice theory.  Recently, Chen and Tong have studied
the D1/D5 effective worldsheet instanton partition function 
on the Higgs branch.  In the gauge theory one looks
at the distribution of instanton size $\rho$
and orientational modes $\hat \Omega$, where
the latter are points on $S^3$.  Indeed, it is
found that the distribution has the $AdS_3 \times S^3$
geometry in the sector with first Chern class $k=1$; that is, a
unit of winding in the $U(1)_{\text{diag}}$ of the color
group.

In a numerical study of this phenomenon, one would
build up a histogram in the $k=1$ topological sector.
Twisted boundary conditions could be imposed to
force nontrivial topology for the gauge fields.
The histogram would count configurations with
a given instanton size $\rho$ and orientation
$\hat \Omega$.  If the weight is identical
to the $AdS_3 \times S^3$ density, it would provide
evidence of the correspondence.  In particular,
it is interesting to explore the correspondence
for intermediate values of $N_c$, given the current
fashion for applying AdS/QCD ideas to real-world
QCD, where $N_c=3$.  

It would also be interesting
to explore the correspondence at finite temperature,
since continuum methods start to break down if
the temperature is too far from zero.  
The recent results of Rey and Hikida 
for small 't Hooft coupling and
finite temperature~\cite{Rey:2006bz} provide continuum
results that could be compared to.
Finally, one would like to study correlation functions
that are not BPS saturated.  Again, continuum
methods are generally unreliable in that case.

\mys{Conclusions}
\label{conc}
In this article I have generalized the EK construction
to 2d (4,4) gauge theories.  I have specialized to
a $U(N_c) \times U(N_f)^n$ quiver theory.  Next,
I showed how to treat the $U(N_f)^n$ quiver as a deconstructed
third dimension and how to obtain a weakly coupled
2d remnant $U(N_f)_{\text{diag}}$, mimicking what really happens in
the D1/D5 brane system.  I described a
simple test of $AdS_3/CFT_2$ that could be
conducted numerically.  It is worth noting
that it should be straightforward
to include Fayet-Iliopoulos (FI) terms in the
mother theory, and thus in the lattice theory;
indeed, this has already been illustrated
by EK in their ``Example 2.''

Work in progress includes a careful study of
renormalization in the lattice theory, the
number of counterterms that need to be fine-tuned,
their exact calculation in perturbation theory
(the lattice theory is super-renormalizable
since the coupling has positive mass dimension),
and a numerical study of the correspondence.
Renormalization of the theory, such as has been
studied in \cite{Onogi:2005cz}, is certainly
a pressing question in the presence of matter.
It remains to be seen the extent to which
complex phase problems of the pure gauge lattice
theory \cite{Giedt:2003vy,Giedt:2004tn}
persist once matter is introduced.
If FI terms are introduced and the theory
is studied on the higgs branch, the complex
phase may be less of a problem.

Finally, it is of some interest to work
out a superfield description of the daughter
theory in this model.  This would be useful
in a super-Feynman diagram perturbative analysis,
as well as for understanding the
renormalizations to the tree-level action
that are possible.

\vspace{15pt}

\noindent
{\bf \Large Acknowledgements}

\vspace{5pt}

\noindent
This work was supported in part by the U.S.~Department of Energy
under grant No.~DE-FG02-94ER-40823.

\myappendix

\mys{Uniqueness of $r_1,r_2$ with conditions imposed}
\label{prfa}
The conditions that we will impose are:
\ben
\item[(i)] The link bosons $z_1,z_2$ should have
$(r_1,r_2) = (1,0)$ and $(0,1)$ respectively.
\item[(ii)] The fermion component $\Psi_4$ should
have $(r_1,r_2) = (0,0)$.
\item[(iii)] At least one other fermion component
in $\Psi,\Psib$ should have $(r_1,r_2)=(0,0)$.
\item[(iv)] All fields should have integral values
of $r_1,r_2$.
\een

It is completely general to write
\beq
r_1 = \sum_{i=1}^4 c_{1i} q_i, \quad
r_2 = \sum_{i=1}^4 c_{2i} q_i.
\eeq
Condition (i) yields immediately
$c_{11}=c_{22}=1, \; c_{21}=c_{12}=0$.
Condition (ii) gives $c_{14}=c_{13}+1,
\; c_{24}=c_{23}+1$.  Thus the
charges reduce to:
\beq
r_1=(q_1+q_4) + c_{13} (q_3+q_4), \quad
r_2=(q_2+q_4) + c_{23} (q_3+q_4).
\eeq
It is easy to see from Table \ref{mtab} that the components
of the matter fermions $\Lambda,\Lamb$ have
$(q_1+q_4) = \pm 1/2, \; (q_2+q_4) = \pm 1/2, \;
(q_3+q_4) = \pm 1/2$.  It follows that
we must take $c_{13}$ and $c_{23}$
to be odd integers, in order to
satisfy condition (iv).  The remaining
condition (iii) then has a unique solution,
as can be checked from Table \ref{vmr}.
It is $c_{13}=c_{23} = -1$, which gives \myref{mmod}.

\mys{Daughter theory matter action}
\label{dmac}
The action can be expressed as three terms,
\beq
S_{\text{mat}} & = & S_1 + S_2 + S_3,
\eeq
where:
\beq
S_1 &=& 
- Q_m^\dagger \( z_{i,m} \zb_{i,m+\ehat_i} 
+ \zb_{i,m} z_{i,m-\ehat_i} \) Q_m \nnn
&& - \Qt_m \( z_{i,m} \zb_{i,m+\ehat_i} 
+ \zb_{i,m} z_{i,m-\ehat_i} \) \Qt_m^\dagger \nnn
&& + F_m^\dagger F_{m+\ihat+\jhat} + \Ft_m \Ft_{m-\ihat-\jhat}^\dagger
+ Q_m^\dagger D_m Q_m - \Qt_m D_m \Qt_m^\dagger \nnn
&& + \sqtw \Big( \Ft_m \phi_{m-\ihat-\jhat} Q_m + \Qt_m G_m Q_m
+ \Qt_m \phi_m F_{m+\ihat+\jhat} \nnn
&& + Q_m^\dagger \phi_m^\dagger \Ft_{m-\ihat-\jhat}^\dagger
+ Q_m^\dagger G_m^\dagger \Qt_m^\dagger 
+ F_m^\dagger \phi_{m+\ihat+\jhat}^\dagger \Qt_m^\dagger \Big),
\eeq
\beq
S_2 &=& \sqtw \Big[ \Lamb_{1,m} \( \zb_{1,m-\jhat} \Lambda_{4,m-\ihat-\jhat}
+ z_{2,m-\jhat} \Lambda_{3,m} \) \nnn
&& + \Lamb_{2,m} \( z_{1,m-\ihat} \Lambda_{3,m} 
- \zb_{2,m-\ihat} \Lambda_{4,m-\ihat-\jhat} \) \nnn
&& - \Lamb_{3,m} \( \zb_{1,m+\ihat+\jhat} \Lambda_{2,m+\jhat}
- \zb_{2,m+\ihat+\jhat} \Lambda_{1,m+\ihat} \) \nnn
&& - \Lamb_{4,m} \( z_{1,m} \Lambda_{1,m+\ihat} 
+ z_{2,m} \Lambda_{2,m+\jhat} \) \Big],
\eeq
\beq
S_3 &=& -\sqtw \Big[ \Lamb_{1,m} \phi_{m-\jhat} \Lambda_{1,m+\ihat}
+ \Lamb_{2,m} \phi_{m-\ihat} \Lambda_{2,m+\jhat} \nnn
&& + \Lamb_{3,m} \phi_{m+\ihat+\jhat}^\dagger \Lambda_{3,m}
+ \Lamb_{4,m} \phi_m^\dagger \Lambda_{4,m-\ihat-\jhat} \nnn
&& - \( \Lamb_{1,m} \Psib_{2,m-\jhat} 
- \Lamb_{2,m} \Psib_{1,m-\ihat} \) Q_m \nnn
&& + \Qt_m \( \Psib_{1,m} \Lambda_{1,m+\ihat}
+ \Psib_{2,m} \Lambda_{2,m+\jhat} \) \nnn
&& - Q_m^\dagger \( \Psi_{4,m} \Lambda_{3,m} 
- \Psi_{3,m} \Lambda_{4,m-\ihat-\jhat} \) \nnn
&& + \( \Lamb_{3,m} \Psi_{3,m+\ihat+\jhat} 
+ \Lamb_{4,m} \Psi_{4,m} \) \Qt_m^\dagger \Big] \nnn
&& + i\sqtw \Big[ Q_m^\dagger \( \Psi_{2,m} \Lambda_{1,m+\ihat}
- \Psi_{1,m} \Lambda_{2,m+\jhat} \) \nnn
&& - \( \Lamb_{3,m} \Psib_{4,m+\ihat+\jhat} 
- \Lamb_{4,m} \Psib_{3,m} \) Q_m \nnn
&& - \(\Lamb_{1,m} \Psi_{1,m-\jhat} 
+ \Lamb_{2,m} \Psi_{2,m-\ihat} \) \Qt_m^\dagger \nnn
&& + \Qt_m \( \Psib_{3,m} \Lambda_{3,m} 
+ \Psib_{4,m} \Lambda_{4,m-\ihat-\jhat} \) \Big].
\eeq
Here, the site indices $m$ are implicitly summed.

\end{document}